\documentclass[11pt]{article}
\usepackage{amstex,psfig,amssymb}
\newcommand{\ol}{\setlength{\itemsep}{0pt.}\begin{enumerate}}
\newcommand{\eol}{\end{enumerate}\setlength{\itemsep}{-\parsep}}
\newcommand{\remove}[1]{}

\newtheorem{thm}{Theorem}

\newtheorem{lemma}[thm]{Lemma}

\newcommand{\qed}{\hfill$\Box$}

\newcommand{\ber}{{\begin{eqnarray*}}}
\newcommand{\eer}{{\end{eqnarray*}}}
\newcommand\bysame{\rule[1mm]{1cm}{.025cm}}
\newcommand\nc\newcommand
\nc\bfa{{\bf a}}\nc\bfA{{\bf A}}\nc\cA{{\cal A}}
\nc\bfb{{\bf b}}\nc\bfB{{\bf B}}\nc\cB{{\cal B}}
\nc\bfc{{\bf c}}\nc\bfC{{\bf C}}\nc\cC{{\cal C}}
\nc\bfd{{\bf d}}\nc\bfD{{\bf D}}\nc\cD{{\cal D}}
\nc\bfe{{\bf e}}\nc\bfE{{\bf E}}\nc\cE{{\cal E}}
\nc\bff{{\bf f}}\nc\bfF{{\bf F}}\nc\cF{{\cal F}}
\nc\bfg{{\bf g}}\nc\bfG{{\bf G}}\nc\cG{{\cal G}}
\nc\bfh{{\bf h}}\nc\bfH{{\bf H}}\nc\cH{{\cal H}}
\nc\bfi{{\bf i}}\nc\bfI{{\bf I}}\nc\cI{{\cal I}}
\nc\bfj{{\bf j}}\nc\bfJ{{\bf J}}\nc\cJ{{\cal J}}
\nc\bfk{{\bf k}}\nc\bfK{{\bf K}}\nc\cK{{\cal K}}
\nc\bfl{{\bf l}}\nc\bfL{{\bf L}}\nc\cL{{\cal L}}
\nc\bfm{{\bf m}}\nc\bfM{{\bf M}}\nc\cM{{\cal M}}
\nc\bfn{{\bf n}}\nc\bfN{{\bf N}}\nc\cN{{\cal N}}
\nc\bfo{{\bf o}}\nc\bfO{{\bf O}}\nc\cO{{\cal O}}
\nc\bfp{{\bf p}}\nc\bfP{{\bf P}}\nc\cP{{\cal P}}
\nc\bfq{{\bf q}}\nc\bfQ{{\bf Q}}\nc\cQ{{\cal Q}}
\nc\bfr{{\bf r}}\nc\bfR{{\bf R}}\nc\cR{{\cal R}}
\nc\bfs{{\bf s}}\nc\bfS{{\bf S}}\nc\cS{{\cal S}}
\nc\bft{{\bf t}}\nc\bfT{{\bf T}}\nc\cT{{\cal T}}
\nc\bfu{{\bf u}}\nc\bfU{{\bf U}}\nc\cU{{\cal U}}
\nc\bfv{{\bf v}}\nc\bfV{{\bf V}}\nc\cV{{\cal V}}
\nc\bfw{{\bf w}}\nc\bfW{{\bf W}}\nc\cW{{\cal W}}
\nc\bfx{{\bf x}}\nc\bfX{{\bf Z}}\nc\cX{{\cal X}}
\nc\bfy{{\bf y}}\nc\bfY{{\bf Y}}\nc\cY{{\cal Y}}
\nc\bfz{{\bf z}}\nc\bfZ{{\bf Z}}\nc\cZ{{\cal Z}}

\def\dim{\qopname\relax{no}{dim}}

\newcommand\Proof{\noindent{\sc Proof. }}
\begin{document}

\title       {Quantum Error Detection II: Bounds }

\author      { Alexei Ashikhmin
\thanks {Los Alamos National Laboratory, Group CIC-3, Mail Stop P990, 
Los Alamos, NM 87545.}\and  
Alexander  Barg
\thanks{Bell Laboratories, Lucent Technologies, 
  600 Mountain Avenue 2C-375,
  Murray Hill, NJ 07974.}\and 
Emanuel  Knill
\thanks { Los Alamos National Laboratory 
  Group CIC-3, Mail Stop P990, 
Los Alamos, NM 87545. 
} \and 
Simon Litsyn
\thanks{ Department of Electrical Engineering-Systems, Tel Aviv University, 
Tel Aviv 69978, Israel. 
}
}

\date{}
\maketitle
\begin{abstract}
In Part I of this paper we formulated the problem of error
detection with quantum codes on the completely depolarized
channel and gave an expression for the probability of
undetected error via the weight enumerators of the code.
In this part we show that there exist quantum codes whose
probability of undetected error falls exponentially with
the length of the code and derive bounds on this exponent.
The lower (existence) bound is proved for stabilizer codes
by the counting argument for classical self-orthogonal
quaternary codes. Upper bounds are proved by
linear programming.
First we formulate two linear programming problems that
are convenient for the analysis of specific short codes.
Next we give a relaxed formulation of the problem in terms
of optimization on the cone of polynomials in the Krawtchouk
basis. We present two general solutions of the problem.
Together they give an upper bound on the exponent of undetected
error. The upper and lower asymptotic bounds coincide for
a certain interval of code rates close to 1. 
\end{abstract}
 

\section{Introduction}

In part I of this paper we defined the undetected error event for 
transmission with quantum codes over a completely depolarized
channel and explained a way to compute its probability via the
weight enumerators of the code. This part is independent of part I
once we agree on the definitions
of a quantum code, the channel and error correction, and the
weight enumerators.
The main results of part I can
be summarized as follows. Let $Q$ be an $((n,K))$ quantum code,
i.e., a linear $K$-dimensional subspace of 
${\cal H}_n:=({\mathbb C}^2)^{\otimes n}.$
\remove{
$$
{\cal H}_n:=\underbrace{{\mathbb C}^{2}\otimes {\mathbb C}^{2} \otimes \ldots 
\otimes {\mathbb C}^{2} 
}_{n \mbox{ times}}
$$}
Let 
\begin{eqnarray*}
B(x,y)&=&\sum_{i=0}^n B_i x^{n-i} y^i\\
B^\bot(x,y)&=&\sum_{i=0}^n B_i^\bot x^{n-i} y^i
\end{eqnarray*}
be the (Shor-Laflamme) weight polynomials of $Q,$
where the weight distributions $B_i,\,0\le i\le n,$ and
$B^\bot_i,0\le i\le n,$ are given by (I.9) and (I.10), 
respectively{\footnote{References (I.9), Theorem I.6, and so on
point to equations, theorems, etc., in the first part of this paper.}.
Then the probability of undetected error equals
\begin{equation}\label{eq:pue}
P_{ue}(Q,p)=\sum_{i=1}^n(B^\bot_i-B_i)\Big({p\over 3}\Big)^i(1-p)^i,
\end{equation}
where $p$ is the probability for a factor $\tau_i$ in the error operator
$E=\tau_1\otimes\dots\otimes\tau_n$ to be nonidentical (this 
probability does not depend on $i$).
In some situations this formula should contain an additional
constant factor (see Theorem I.6) which throughout this part
will be omitted.

The goal of this part of the paper is to derive bounds on
$P_{ue}(Q,p)$ for the best possible code $Q$ with given
parameters. More specifically, in Part I we defined the quantity
\[
P_{ue}(n,K,p)=\min\limits_{Q\in {\cal H}_n\atop \dim(Q)=K} P_{ue}(Q,p).
\]
We will derive upper and lower bounds on $P_{ue}(n,K,p)$.
Just as in the classical case, this probability falls exponentially
with $n$; therefore, let us also introduce the exponent of undetected error
$$
E(n,R_Q,p)=-{1 \over n}\log_2 P_{ue}(\lceil 2^{R_Qn}\rceil,n,p),
$$
where $R_Q={\log_2K\over n}$ is the code rate.
Speaking of asymptotics, we are interested in upper bounds 
$$
\overline{E}(R_Q,p)=\limsup_{n\to \infty} E(n,R_Q,p)
$$
and lower bounds 
$$
\underline{E}(R_Q,p)=\liminf_{n\to \infty} E(n,R_Q,p)
$$
(this corresponds to lower and upper bounds on the probability 
$P_{ue}(n,K,p),$ respectively).
Let $E(R_Q,p)$ be the common limit of these functions, provided that it 
exists. 

Throughout the paper 
\begin{align*}
T_q(x,y)&=x\log_q(q-1)-x\log_qy-(1-x)\log_q(1-y)\\
H_q(x)&=T_q(x,x).
\end{align*}

In the classical case,  error detection has been
extensively studied. The probability of undetected error in the 
classical case is defined
in (I.1); its exponent $E^{(cl)}(R,p)$ is defined similarly to the above.
Best known lower bounds on $E^{(cl)}(R,p)$ (upper bounds on the probability)  
were derived in \cite{Levenshtein1} building upon the 
Varshamov-Gilbert-type existence arguments. We consider the
binary case only. Let $R_{vg}(x):=1-H_2(x)$ be the 
Varshamov-Gilbert function and $\delta_{vg}(x)$ its inverse.
Also let $\bar R (x)$ be the function giving the best known
upper bound on codes \cite{ref mcel} and $\bar \delta(x)$ its inverse.
It is easy to 
prove that there exist binary linear codes with 
$A_i\le n{n\choose i}2^{k-n},$
where $A_i$ is the number of vectors of weight $i$ in the code.
Substituting this in (I.1), we obtain the lower bound 
of \cite{Levenshtein1}:
\begin{equation*}
E^{(cl)}(R,p)\ge\begin{cases} T_2(\delta_{vg}(R),p) &0\le R\le 1-H(p)\\
1-R &1-H_2(p)\le R\le 1.
\end{cases}
\end{equation*}
Upper bounds require more involved arguments \cite{ab},\,\cite{lit1}.
The results have the form
\begin{equation*}
E^{(cl)}(R,p)\le\begin{cases} 
1-R-H_2(\bar\delta(R),p)+T_2(\bar\delta(R),p) &0\le R\le \bar R(p)\\
1-R &\bar R(p)\le R\le 1.
\end{cases}
\end{equation*}
In this paper we derive analogous results for the quantum case.
In Section \ref{upperb} we prove the existence of quantum 
stabilizer codes with bounded above weight enumerators; a substitution
in (\ref{eq:pue}) yields lower bounds on $E(R,p).$
In this part we rely upon the results of \cite{macw78} on
quaternary self-orthogonal codes.
Then we move on to lower bounds on $P_{ue}(n,K,p)$.
In this part we employ the linear programming technique.
In Section \ref{lpprob}  we formulate a linear programming (LP) problem 
with the objective function $P_{ue}(n,K,p)$.
Though in examples this problem gives good lower bounds
(which can be found by solving it on a computer), analytically
it is difficult to deal with.
Therefore, in the second part of this section we propose a relaxation
of the problem which enables us to derive general bounds. 
This part of the paper is based on an application of the
LP approach in the quantum case \cite{al} in conjunction
with the methods of \cite{ab},\,\cite{lit1}.
The results include two upper bounds on $E(R,p)$.
These bounds show that for the rate $R_Q$ in a certain
neighborhood of 1, dependent on p, the exponent $E(R,p)$
is known exactly. For lower rates, the bounds are in general
location, i.e., there exists a segment in which 
each of them is better than the other.
This part of the paper is technically the most involved.
We chose to formulate the results for arbitrary $q$
(instead of concentrating on $q=4$), the reasons being
that once we look at $q>2,$ it does not make much of a difference
whether it is $4$ or anything else, and that this is helpful
in studying error detection of nonbinary classical codes
on which we plan to report elsewhere. Moreover, the theory
of quantum codes generalizes to larger $q$
\cite{knill96},\,\cite{rains}, 
though the presentation is somewhat less systematic and 
the results more scattered than for binary quantum codes.

Some further remarks on notation. Throughout the paper 
$F={\mathbb F}_4=\{0,1,\omega,\omega^2\}.$
The Krawtchouk polynomial is given by
\[
K_k(q;x)=\sum_{\ell=0}^n(-1)^\ell{x\choose \ell}{n-x\choose k-\ell}
(q-1)^{k-\ell}
\]
(the implicit parameter $n$ -- the length of the code -- is usually
clear from the context). Properties of $K_k(q;x)$ used
throughout the paper are summarized in the appendix.
As remarked above, by $R_Q$ we denote the
rate of the quantum code. We also use two associated numbers
$|C|=\sum_{i=0}^n B_i$ and $|C^\bot|=\sum_{i=0}^n B_i^\bot;$
in the case of stabilizer codes they are equal to the size of 
the two underlying classical codes, $C$ and $C^\bot$ (see Part I).
Likewise, let $R=(1/n)\log_4|C|,$ $R^\bot=(1/n)\log_4|C^\bot|.$
The rate $R_Q$ and these 2 quantities are connected by the following 
relations:
\begin{equation}\label{eq:r}
R_Q=2R^\bot-1=1-2R, \quad R+R^\bot=1,\quad 0\le R\le{1\over2}\le
R^\bot\le1.
\end{equation}

\remove{{\sc Summary of results.} The main results of this paper are
given in the following theorem.
\begin{thm} \label{thm:main}
\newline
{\rm(A)} 
\begin{equation}\label{eq:bounds-lower}
E(R,p)\le
v\begin{cases}
{1 \over 2} - {R\over 2},   & {1\over 2}+{R\over 2}\ge 1-H_4(p)\\
T_4\Big(H_4^{-1}\Big( {1 \over 2} - {R_Q \over 2} \Big),p\Big)  
&\mbox{otherwise}.
\end{cases}
\begin{tabular}{cc}{\rm (I)}\\[2mm]{\rm (II)}\end{tabular}
\end{equation}
Moreover, there exist sequences of quantum stabilizer codes
that attain these bounds.

(B) \begin{equation}\label{eq:bounds-upper}
E(R,p)\ge
\begin{cases}
\end{cases}
\end{equation}
\end{thm}
}
 
\section{Upper (existence) bounds on $P_{ue}(n,K,p)$}
\label{upperb}

In this section we show that there exist quantum codes
for which probability of undetected error falls exponentially
for all rates $0\le R_Q<1,$ and bound this exponent below
More specifically, we prove the following theorem.
\begin{thm}
\label{GV} 
\begin{equation*}
E(R_Q,p)\ge 
\begin{cases}
T_4\Big(H_4^{-1}\big( {1 \over 2}(1 - {R_Q})\big),p\Big)  &
0\le R_Q\le  2(1-H_4(p))-1;\\
{1 \over 2}(1- {R_Q})   & 2(1-H_4(p))-1\le R_Q\le 1. 
\end{cases}
\end{equation*} 
\end{thm}

To prove this theorem, we restrict our attention to quantum
stabilizer codes. In analogy with the classical case, we show
that there exist sequences of codes $Q$ of growing length $n$ and
size $K=2^{k_Q}$ with weight distribution
$$
B_i^\bot\le n^2{{n \choose i}3^i 2^{k_Q-n}}.
$$ 
(in fact, $n^2$ can be easily replaced by $n$).

Since the weight distributions $B_i, B_i^\bot$ correspond
to classical quaternary code, we prove this estimate by considering 
families of quaternary self-orthogonal codes. Let $C\subset F^n$.
Throughout the section we denote by $C^\bot$ a linear code dual to $C$
with respect to the standard dot product $(\cdot,\cdot).$
Let
\begin{align*}
\cS_{n,k}&=\{C\subset F\mid C \mbox{ even linear code}, \dim C=k\} \\
\cS_{n,k}^\bot&=\{C\subset F\mid C^\bot\in \cS_{n,k}\},
\end{align*}
where $k\le n/2$ by (\ref{eq:r}).

\remove{
\noindent {\bf Remark} In fact we will show even stronger result that there 
exists a quantum stabilizer code
that has binomial distribution for both $B_i$ and $B_i^\bot$, i.e. 
$$
B_i=\Big \lceil {1 \over 2^{n+k}} 3^i {n \choose i} \Big \rceil,  
B_i^\bot=\Big \lceil {1  
\over 2^{n-k}} 3^i {n \choose i} \Big \rceil.
$$}
Below we use the following three results from \cite{macw78}. 

\begin{lemma} 
\label{even} Let $C\subset F^n$ be an even linear code. Then $C$ is 
self-orthogonal with respect to
the inner product $\bfa\ast\bfb=\sum_{i=1}^n(a_ib_i^2+a_i^2b_i)$.
\end{lemma} 
 
\begin{lemma} 
\label{num_of_even} Let $C\subset F^n$ be an even linear code
and $C^\bot$ be its dual.
Then the number of even-weight code vectors in
$C^\bot$ equals
\(
{1\over 2}( 4^{n-k}+(-2)^n). 
\)
\end{lemma} 

\begin{lemma} 
\label{even_coset} Let $C\subset F^n$ be an even linear code
and $C^\bot$ be its dual. If $\bfa\in C^\bot$ has even (odd) weight
then the coset $C+\bfa$ is formed by vectors of even (odd)
weight.
\end{lemma}

Existence of codes with bounded distance distribution will
follow from the following lemma, based on Lemmas \ref{even}-\ref{even_coset}.
\begin{lemma}
\label{the_same} Let $\bfv\in F^n$ be any even-weight vector. The number
of codes from $\cS$ containing $\bfv$ does not depend on $\bfv$.
\end{lemma}
\Proof
Let us count the number of linear $[n,k]$ codes containing $\bfv.$ 
Let $C_1=\langle\bfv\rangle$ be the $[n,1]$ code
and $\bfw$ an even-weight vector distinct from $\bfv$ such that 
$(\bfw,\bfv)=0$. By Lemma \ref{even_coset} all the cosets 
$C_1+\alpha\bfw, \alpha\in F,$ are even; so by adjoining $\bfw$
we obtain an even $[n,2]$ code $C_2$. By 
Lemma \ref{num_of_even} this can be done in 
$$
{1\over 4} \Big[{1\over 2}\Big(4^{n-1}+(-2)^n \Big)-1\Big]
$$
ways independent of ${\bf v}$. Similarly $C_2$ can be extended to an even 
$[n,3]$ code in
$$
{1\over 16}\Big[ {1\over 2}\Big(4^{n-2}+(-2)^n \Big)-1\Big]
$$
ways. Continuing in this manner, we obtain all even $[n,k]$ codes 
that contain ${\bf v}$. 
It is obvious that their number does not depend on a particular 
choice of ${\bf v}$.
\qed

\begin{thm}
\label{s}
The family $\cS_{n,k}$ of even $[n,k]$ codes contains a code $C$
with weight distribution
\(
B_i(C)\le n^2 \tilde B_i,\,1\le i\le n,
\)
where 
\[
\tilde B_i={ 4^k-1 \over {1\over 2}\Big(4^{n-1}+(-2)^n \Big)-1 } 
{n \choose i} 3^i
\]
is the average weight distribution of codes in $\cS_{n,k}.$
\end{thm}
\Proof Let $N=|\cS_{n,k}|.$
By Lemma \ref{the_same} every even vector $\bfv$ is contained
in one and the same number, say $L$, of codes from $\cS_{n,k}.$
So computing the total number of all vectors in codes 
from $\cS_{n,k}$ in two ways, we get
$$
N+\Big(  {1\over 2}\Big(4^{n-1}+(-2)^n \Big)-1 \Big)L=4^kN,  
$$
or
$$ 
N=L { {1\over 2}\Big(4^{n-1}+(-2)^n \Big)-1 
\over 4^k-1 }. 
$$ 
Let $B_i(C_j)$ be the number of code vectors of weight $i$ in the $j$-th 
code from $\cS_{n,k}$. Then we have
$$
\sum_{j=1}^N B_i(j)={n \choose i} 3^i L. 
$$
Hence the average over $\cS_{n,k}$ number of codewords of weight $i$ is 
${L\over N}{n \choose i} 3^i=\tilde B_i,$ as claimed. 
The number of  codes $C\in \cS_{n,k}$ such that $B_i(C)\ge n^2 \tilde B_i$
for a given $i$ is not greater than 
\[
{ \sum_{j=1}^N B_i(C_j) \over n^2  \tilde B_i }={L \over n^2}  
{ {1\over 2}\Big(4^{n-1}+(-2)^n \Big)-1 \over 4^k-1} = {1 \over n^2} N. 
\]
Hence the number of codes $C$ such that $B_i(C)\le n^2  \tilde B_i $ for all 
$1\le i\le n$ is at least  
\[
N-{N \over n}=N\Big( 1- {1 \over n} \Big).
\]
\remove{
To finish the proof we note that
\[
{1 \over n}\log n^2  \tilde B_i = R-1+H_4\Big( {i \over n} \Big) +o(1).
\]}
\vskip-5mm\qed


Now let us use this result to prove that
the family $\cS^\bot_{n,k}$ also contains codes
whose weight distribution is bounded above by a polynomial
factor times the average weight distribution in $\cS^\bot_{n,k}$. 
This will enable us to
prove Theorem \ref{GV}.
In this part we rely on the MacWilliams identities.
We will need the following lemma.
\begin{lemma}
\label{2i}
Let $n$ be an even integer. Then
\[
\sum_{i=0}^{n/2} {n \choose 2i} 3^{2i} K_t(4;2i)=2^{n-1} {n \choose t}(-3)^t.
\]
\end{lemma}
\Proof By (\ref{gen_fun}), the sum 
$\sum_{i=0}^{n/2}{n \choose 2i}3^{2i} K_r(4;2i)K_s(4;2i)$ is the 
coefficient of $y^rz^s$ in 
\begin{align}
\sum_{i=0}^{n/2} {n \choose 2i} 
3^{2i}&(1+3y)^{n-2i}(1-y)^{2i}(1+3z)^{n-2i}(1-z)^{2i} \nonumber \\
&={1 \over 2} \Big( [ (1+3y)(1+3z)+3(1-y)(1-z)]^n \nonumber\\
&\phantom{={1 \over 2}}+[(1+3y)(1+3z)-3(1-y)(1-z)]^n\Big) \nonumber \\
&= {1 \over 2} \Big( [4+12yz]^n+2^n[-1+3(y+z)+3yz]^n\Big). \label{genf}
\end{align}
It is clear that the first term in (\ref{genf}) contributes only to 
coefficients of $y^rz^r$. Consider the second term:
\begin{gather}
2^n[-1+3(y+z)+3yz]^n
= 2^n \sum_{i=0}^n {n \choose i} 3^{n-i}y^{n-i}z^{n-i}  
(3(y+z)-1)^i\label{genf1}
\end{gather}
Since we are interested only in the coefficient  $y^0z^t$,  
in the sum (\ref{genf1}) we put $i=n$. This gives 
\begin{eqnarray*}
2^n (3(y+z)-1)^n
&= & 2^n \sum_{i=0}^n {n \choose i} (-3)^i \sum_{j=0}^i {i \choose j} y^j 
z^{i-j}  \\
& = & 2^n \sum_{j=0}^n y^j \sum_{i=0}^n {n \choose i} {i \choose j} (-3)^i 
z^{i-j}
\end{eqnarray*}
The coefficient of $y^0z^t$ in this sum equals 
$
2^n {n \choose t} {(-3)^t}.
$
\qed

\begin{thm}
\label{sperp}
In $\cS^\bot_{n,k}$ there exists a code $C$ with
\(
B_i^\bot\le n^2 \tilde B_i^\bot, \,1\le i\le n,
\)
where 
\begin{equation}\label{eq:bperp}
\tilde B_i^\bot={1 \over 4^k} {n 
\choose i} 3^i \Big(1+ {((-1)^i2^{n-1}
-1)(4^k-1) \over {1\over 2}(4^{n-1}+(-2)^n )-1} \Big)
\end{equation}
is the average weight distribution of codes in $\cS^\bot_{n,k}$.
\end{thm}
\Proof As in Theorem \ref{s}, it suffices to compute the
average. Using the MacWilliams identities and the fact that
$|\cS^\bot_{n,k}|=|\cS_{n,k}|,$ we have, for a given $t,$
\begin{eqnarray*}
\sum_{j=1}^N B_t^\bot (C_j^\perp)& = &{1 \over 4^k} \sum_{j=1}^N 
\sum_{i=0}^n B_i(j) 
K_t(4;i)\\
& = &{1 \over 4^k} \sum_{i=0}^n \Big( \sum_{j=1}^N B_i(j) \Big) K_t(4;i) \\
 &= &{1 \over 4^k}\Big( NK_t(4;0)+L\sum_{i=0}^{n/2} {n \choose 2i} 
3^{2i} K_t(4;2i) 
-LK_t(4;0)\Big). 
\end{eqnarray*}
  From Lemma \ref{2i} it follows that for $t>0$ 
\begin{eqnarray}
vv\sum_{j=1}^N B_t^\bot (C_j) &= &{1 \over 4^k} \Big( N{n \choose t}3^t+L 
2^{n-1} 
(-3)^t{n \choose t}-L{n \choose t}3^t\Big ) 
\nonumber \\
&=& {1 \over 4^k} N {n \choose t} 3^t \Big[ 1+ ((-1)^t2^{n-1}-1)
{4^k-1 \over {1\over 2}(4^{n-1}+(-2)^n )-1} \Big].\nonumber
\end{eqnarray}
Hence
$$
\tilde B_t^\bot= {1 \over N} \sum_{j=1}^N B_t^\bot (C_j).
$$
The proof is completed as in Theorem \ref{s}. \qed
\remove{
Noting that $\lim_{n\to \infty } {((-1)^t2^{n-1}-1)(4^k-1) \over {1\over 
2}\Big(4^{n-1}+(-2)^n \Big)-1}=0,$ 
when $k=\lambda n, \lambda<0.5$, 
we have ${1 \over n} \log \tilde B_t=H_4(t/n)-1+R^\bot+o(1)$. 
Now using the 
same arguments as in Theorem \ref{s}, we get that
at least $N\Big(1-{1 \over n} \Big)$ codes from $\cS^\bot$ have 
$$
B^\bot_i\le n^2 \tilde B_i^\bot, i=0,1,\ldots ,n. 
$$
\qed }

Now we are in a position to prove Theorem \ref{GV}. Let $Q$ be a 
quantum code satisfying Theorems \ref{s} and \ref{sperp}.
Note that the second term in the expression (\ref{eq:bperp})
for $\tilde B_i^\bot$ vanishes as $n$ grows; so starting with
some value of $n$ the weight distribution $B_i^\bot$ is bounded
above as
\begin{equation}\label{eq:binomial}
B_i^\bot\le 2n^2{n\choose i}3^i2^{k_Q-n}.
\end{equation}
Let us compute $P_{ue}(Q,p)$ for large $n$ relying on this inequality.
We have
\begin{align}
P_{ue}(Q,p) &=\sum_{i=0}^n (B_i^\bot -B_i) \Big({p \over 3}\Big)^i 
(1-p)^{n-i}\nonumber\\
&\le \sum_{i=1}^n B_i^\bot  \Big({p \over 3}\Big)^i (1-p)^{n-i}\nonumber\\
&\le n^2
\sum_{i=1}^n 2^{k_Q-n+1} {n\choose i}3^i\Big({p \over 3}\Big)^i (1-p)^{n-i}
\label{eq:sum}
\end{align}
The exponent of the summation term equals
\begin{equation}\label{eq:exp1}
-n\Big[H_4\Big({i\over n}\Big)-{i\over n}\log_4{p}
-\Big(1-{i\over n}\Big)\log_4(1-p)+{1\over2}(1-R_Q)\Big],
\end{equation}
where we have omitted the $o(1)$ terms. The expression in brackets
attains its maximum for $(i/n)=p.$ 
Note also that the right-hand side in (\ref{eq:binomial}) behaves 
exponentially in $n$; the exponent approaches $0$ when
the quotient $(i/n)\to \delta_{vg}(R^\bot)$. Hence as long as
$(i/n)\ge \delta_{vg}(R^\bot),$ the sum in (\ref{eq:sum})
asymptotically is dominated by the term with $i=\lfloor np\rfloor$.
Thus as long as $1-H_4(p)\le R^\bot={1\over2}(1+R_Q),$
we have
\[
E(R_Q,p)\le {1\over 2}(1-R_Q),
\]
i.e., the second part of the theorem.
Otherwise, the maximum moves outside the summation range,
so the largest term asymptotically is the first one in the sum, i.e.,
the one corresponding to $i=\lfloor n \delta_{vg}(R^\bot)\rfloor$. 
This gives the first part of the theorem. \qed

Note that we have proved a stronger fact about quantum codes
than the one actually in Theorem \ref{GV}, namely, that there
exist stabilizer codes both of whose weight distributions
$B_i$ and $B_i^\bot$ are bounded above by the ``binomial'' term
$n^2{n\choose i}3^{i}2^{k_Q-n}$.
 
\section{A linear program for quantum undetected error}
 \label{lpprob}

The approach leading to best known lower bounds on the probability
of undetected error in the classical case has been the 
linear programming one \cite{ab},\,\cite{lit1}.
In this section we develop a similar technique for the quantum case.
First we formulate two theorems that enable one to obtain
good lower estimates on $P_{ue}(n,K,p)$ for finite $n$.
Then we formulate a relaxed LP problem which is not as good for finite
$n$ but lends itself to asymptotic analysis.

For the reasons outlined in the introduction
we will study a general alphabet of size $r.$
An $r$-ary quantum code $Q$ is a $K$-dimensional linear subspace
of ${\mathbb C}^{r^n}.$ Without going into details we say that one 
can associate with $Q$ two weight distributions,
$B_i$ and $B_i^\bot,\,0\le i\le n,$ connected by the $q$-ary
MacWilliams identities, $q=r^2.$ Furthermore, 
\[
K={1\over q^{n}}\sum_{i=0}^n B_i^\bot.
\]
As above, we use the notation 
\begin{gather*}
|C^\bot|=\sum_{i=0}^n B_i^\bot, \quad |C|=\sum_{i+0}^n B_i,\\
R=(1/n)\log_q|C|,\quad  R^\bot=(1/n)\log_q|C^\bot|;
\end{gather*}
these numbers and the rate $R_Q$ are again related through (\ref{eq:r}).
We have for the probability of undetected error
\begin{equation}\label{qpue}
P_{ue}(Q,p)=\sum_{j=1}^n (B_j^\bot-B_j) \Big({p \over q-1}\Big)^j 
(1-p)^{n-j}.
\end{equation}
Our first result is given by the following theorem.

\begin{thm}
\label{ajperp} Let $Q$ be an $((n,r^{nR_Q}))$ quantum code.
Let $q=r^2,$ $R^\bot=(1+R_Q)/2.$
Let $Z(x)=\sum_{i=0}^n z_i K_i(q;x)$ and 
$Y(x)=\sum_{i=0}^n y_iK_i(q,x)$ be polynomials such that
\begin{align}
Z(j)-& Y(j)+y_0+y_j q^{n{R^\bot}} \nonumber \\
&\le q^{n{R^\bot}} \Big( {p \over q-1}\Big)^j (1-p)^{n-j}
-\Big( {q-1-qp \over q-1}\Big)^j, 1\le j \le n,  
\label{pol_rest} \\
z_0,y_0 \lesseqqgtr &0;\;z_j\ge 0,\; y_j\ge 0, 1\le j \le n. \nonumber
\end{align}
Then
\[
P_{ue}(n,r^{nR_Q},p)\ge q^{-n{R^\bot}} \Big(z_0q^{n{R^\bot}}-Z(0)+Y(0)-y_0\Big)
+(1-p)^n-q^{-nR^\bot}.
\]
\end{thm}
\Proof
Using the MacWilliams identities we can rewrite (\ref{qpue}) as follows:
\begin{eqnarray*}
P_{ue}(Q,p) &= &{1 \over |C^\bot |} \sum_{j=1}^n B_j^\bot \Big[ |C^\bot | 
v\Big( {p \over q-1 }
\Big)^j (1-p)^{n-j}\\
&& -\sum_{t=1}^n K_t(j)\Big({p \over q-1 }\Big)^t (1-p)^{n-t} \Big]
-{1 \over |C^\bot | } \sum_{j=1}^n {n \choose j} p^j (1-p)^{n-j}.
\end{eqnarray*}
The middle term here is calculated using the generating function 
(\ref{gen_fun}):
\[
\sum_{t=0}^nK_t(j)\Big( {p \over q-1}\Big)^t (1-p)^{n-t}
=\Big({q-1-qp \over q-1 }\Big)^j.
\]
Hence
\begin{eqnarray}
\label{podg}
P_{ue}(Q, p) & =  & {1 \over |C^\bot |} \sum_{j=1}^n B_j^\bot 
\Big[ |C^\bot 
|\Big( {p \over q-1 } \Big)^j
(1-p)^{n-j} - \Big( {q-1-qp \over q-1} \Big)^j +(1-p)^n \Big]  
\nonumber \\
&  &- {1 \over |C^\bot |} \sum_{j=1}^n 
{n \choose j} p^j (1-p)^{n-j}  \nonumber \\
& = & {1 \over |C^\bot |} \sum_{j=1}^n B_j^\bot \Big[ |C^\bot |
\Big( {p 
\over q-1 } \Big)^j
(1-p)^{n-j} - \Big( {q-1-qp \over q-1 }\Big)^j\Big]\nonumber \\
&  &+(1-p)^n-{1\over |C^\bot|}.
\end{eqnarray}

Thus, $P_{ue}(Q, p)$ is a linear form of the coefficients
$B_i^\bot$ which we have to minimize. We can formulate the
following LP problem:
\begin{equation}
\label{prlp}
\min\Big\{\sum_{j=1}^n B_j^\bot \Big[ |C^\bot |\Big( {p \over q-1 } \Big)^j
(1-p)^{n-j} - \Big( {q-1-qp \over q-1} \Big)^j   \Big]\Big\}
\end{equation}
subject to the restrictions
\begin{eqnarray*}
B_j^\bot & \ge & 0,\quad 1\le j \le n \\
\sum_{j=1}^n B_j^\bot &=& |C^\bot |-1 \\
\sum_{j=1}^n B_j^\bot K_i(q;j) &\ge & -{n \choose i} (q-1)^i,\quad
 1\le i \le n\\
B_j^\bot |C^\bot |-\sum_{i=1}^n B_i^\bot K_j(q;i) &\ge & 
{n \choose j} (q-1)^j, \quad 1\le j \le n.
\end{eqnarray*}
The last inequality follows from Theorem I.2(i).

Now the theorem follows by the LP duality. Indeed, the dual
program has $2n+1$ variables $(z_0, z_1, \ldots ,z_n)$ and 
$(y_1, y_2, \ldots ,y_n)$ 
of which $z_0$ can take on any value and all the other variables are 
nonnegative. The dual objective function has the form
\[
\max\Big\{z_0(|C^\bot|-1)-\sum_{i=1}^n (z_i-y_i) {n \choose i}(q-1)^i \Big\}
\]
subject to restrictions
\remove{\begin{equation}
\label{rest1}
z_0 \mbox{ is any,  }
z_j\ge 0, y_j\ge 0, 1\le j \le n
\end{equation}}
\begin{align*}
z_0+&\sum_{i=1}^n  (z_i-y_i) K_i(q;j)+y_j |C^\bot |\\
 &\le  |C^\bot| \Big( {p\over q-1}\Big)^j (1-p)^{n-j} -
\Big( {q-1-qp \over 
q-1}\Big)^j\quad 1\le j\le n. 
\end{align*}
 
Any feasible solution of the this problem gives a lower estimate of
$P_{ue}(n,K,p)$. Let us introduce the polynomials
$Y(x)=\sum_{i=0}^ny_iK_i(q;x)$ and $Z(x)=\sum_{i=0}z_i K_i(q;x).$
Since $K_0(q;x)\equiv1,$ this implies our claim. \qed

Sometimes it is convenient to rewrite the linear problem via the 
enumerator  $B_j$ instead of $B_j^\bot$. The proof of the
following theorem is similar to the above.
\begin{thm}
\label{aj} Suppose $Q, n,r,R_Q, q, Z(x),Y(x)$ have the
same meaning as in Theorem {\rm \ref{ajperp}}. Let $R=(1-R_Q)/2.$
Suppose that
\begin{align}
Z(j)+&Y(j)-y_0-y_jq^{nR}\nonumber \\
& \le \Big( {q-1-pq \over q-1}\Big)^j- q^{nR} 
\Big( {p \over 
q-1}\Big)^j (1-p)^{n-j},\quad
 1\le j \le n, \label{pol_rest_2}\\
z_0,y_0\lesseqqgtr &0,\;z_j\ge 0,\; y_j\ge 0,\quad 1\le j \le n.\nonumber
\end{align}
Then
\[
P_{ue}(R_q,p)\ge q^{-nR} \Big(z_0q^{nR}-Z(0)-Y(0)+y_0\Big)-(1-p)^n+
q^{-nR}.
\]
\end{thm}

Though the LP problems of Theorems \ref{ajperp} and \ref{aj} enable one 
to find bounds for short codes with the use of computer,
they are not easy to analyze in general. 
The reason for this is that the sign of the quantities on the 
right-hand side of (\ref{pol_rest}) or (\ref{pol_rest_2})
alternates. This significantly complicates checking feasibility of a 
putative solution. For this reason below we take on a different
approach which, though it does not yield optimal solutions for the LP
problem, gives rise to good asymptotic upper bounds on $E(R,p).$

\begin{thm}
\label{key}
Let $Q$ be an $r$-ary quantum code with weight enumerators $B_i$ and 
$B_i^\bot.$ Let $h(i),1\le i\le n,$ be a real-valued function and  
$$
Z(x)=\sum_{i=0}^n z_i K_i(q;x),\quad(q=r^2)
$$
be a polynomial  that for $1\le i\le n$ satisfies the conditions
\begin{align}
&\mbox{\rm(i)}\quad Z(i)\le h(i); \label{eq:ii}\\
&\mbox{\rm(ii)}\quad z_i|C^\bot|-Z(i)\ge 0. \label{eq:iii}
\end{align}
Then
\begin{equation}\label{eq:main-bound}
\sum_{i=0}^n (B_i^\bot-B_i) h(i)\ge z_0|C^\bot|-Z(0).
\end{equation}
\end{thm}
\Proof The proof will follow from the following chain of
relations:
\begin{align*}
\sum_{i=0}^n (B_i^\bot-B_i) h(i)&\ge \sum_{i=0}^n (B_i^\bot-B_i) Z(i)\\
&\stackrel{\mbox{\footnotesize(a)}}=
\sum_{i=0}^n\Big( {1 \over |C|} \sum_{j=0}^n B_jK_i(q;j)-B_i 
\Big) Z(i)\\
&=\sum_{j=0}^n B_j  {1 \over |C|}  \sum_{i=0}^n K_i(q;j)Z(i)-
\sum_{i=0}^n B_i Z(i)\\
&\stackrel{\mbox{\footnotesize(b)}}=
\sum_{j=0}^n B_j {r^n \over |C|} z_j - \sum_{j=0}^n B_j Z(j) 
=\sum_{j=0}^n B_j (|C^\bot | z_j -Z(j))\\
&\ge |C^\bot |z_0-Z(0),
\end{align*} 
where the first inequality follows by (i) and the obvious
$B_0=B^\bot_0=1$; step (a) is implied
by the MacWilliams identities, in (b) we use (\ref{coef}), 
and the final inequality follows by (ii) and the fact that
$B_j\ge 0, B_0=1.$ \qed

If (\ref{eq:iii}) is replaced by the condition 
\begin{equation}\label{eq:iiii}
Z(i)-|C|z_i\ge 0,\,i=1,\dots,n,
\end{equation}
then 
by a similar argument one can prove that
\[
\sum_{i=0}^n (B_i^\bot-B_i) h(i)\ge Z(0)- |C|z_0.
\] 

We wish to stress the difference between the conditions on $Z(x)$
in this theorem and in the classical (non-quantum) case \cite{ab}.
The standard condition $z_j\ge 0, 1\le j\le n,$ is not needed to prove
(\ref{eq:main-bound}); it is replaced by related though not
equivalent conditions (\ref{eq:iii}), (\ref{eq:iiii}). 
In the situation when one bounds above the size of the quantum code, 
the corresponding inequality is $|C|\le \max_{1\le j\le n}(Z(j)/z_j);$
see \cite{al} for details.

In the next section we use Theorem \ref{key} 
to derive asymptotic bounds on $E(R,p).$

\section{Lower Bounds on $P_{ue}(n,K,p)$}
\label{lpbounds}

In this section we prove two lower bounds on the probability
of undetected error that are valid for any quantum code
of given length and size. The bounds are derived by a suitable
choice of polynomials in Theorem \ref{key}. The results are
similar in spirit to \cite{ab},\,\cite{al},\,\cite{lit1}.
In this section $\gamma=q-1,$
\begin{equation}
h(x)=({p/\gamma})^{x}(1-p)^{n-x}\label{eq:h(x)},
\end{equation}
and as usual $0\le p< \gamma/q.$

\subsection{An Aaltonen-MRRW-type bound }
In this part we rely on the technique in \cite{ref mcel}\footnote{The
abbreviation in the title is derived from its authors' names.}, extended
to arbitrary $q$ in \cite{aal77}, and apply it in a way similar
to \cite{ab},\,\cite{al}. Let
\begin{equation}\label{eq:tau}
\tau_0(z):={\gamma\over q}-{\gamma-1\over q}z-
{2\over q}\sqrt{\gamma z(1-z)} \quad(0\le z\le (\gamma/q)).
\end{equation}
By \cite{ref mcel},\,\cite{aal77} $R_q^{lp}(\delta):=H_q(\tau_0(\delta))$
is the maximal asymptotically attainable rate of a $q$-ary (classical) 
code with relative distance $\delta.$ 
This is proved by studying Delsarte's linear programming problem
with the polynomial
\[
f_t(x)={1 \over (a-x)} (K_t(q;x)+K_{t+1}(q;x))^2,
\]
where $t=\lfloor n\tau\rfloor,\, \tau=\tau_0(\delta),$ 
and $a$ is the smallest zero
of $K_t(q;x)+K_{t+1}(q;x).$ Conversely, the function 
$\delta_q^{lp}(R):=\tau_0(H_q^{-1}(R))$
gives an asymptotic upper bound on the
minimum distance of $q$-ary codes of rate $R$. (An interesting remark
is that the function $\tau_0(\cdot)$ is involutive; this is
intimately related to the self-duality of the Hamming scheme 
\cite{del73}). 

We begin with an appropriate modification (rescaling) of the
polynomial $f(x).$  Let 
\begin{equation}\label{pollp1}
Z_t(x)={h(x^\ast)\over K_t^2(q;x^{\ast})} f_t(x),
\end{equation}
where 
$x^{\ast}=\xi^{\ast}n$ is an integer parameter, $0 <x^{\ast}
<n,$ and $t=\lfloor n\tau_0(\xi^\ast)\rfloor$. This choice is motivated
by the following argument. The polynomial has to satisfy the inequality
(\ref{eq:ii}); any reasonable choice of $Z(x)$ 
suggests that it be equal to $h(x)$ at least at one point $x=x^\ast$.
This point is left a free parameter, chosen later. 

The program (\ref{pollp1}) gives rise to the following bound.
For the reasons revealed in \cite{al} and outlined in footnote 3 below,
in the quantum case the bound is valid for all but very low rates.
Below $\alpha_1(q)$ is a certain small positive number dependent
only on $q$. It can be computed for any $q$; for instance, 
$\alpha_1(4)\approx 0.0028.$ We could not find a closed-form
expression for it. 
\begin{thm}
\label{thm:qlp1} Let $\alpha_1(q)\le R_Q\le 1,$
$R=(1/2)(1-R_Q).$ Then 
\begin{equation}
\label{lp1}
E(R_Q,p)\le \begin{cases}
R-H_q(\delta_q^{lp}(1-R))+T_q(\delta_q^{lp}(1-R),p)
& \alpha_1(q)\le R_Q\le 2R_q^{lp}(p)-1;\\
R &2R_q^{lp}(p)-1\le R_Q\le 1.\end{cases}
\end{equation}
\end{thm}
\Proof
We first prove the bound (\ref{lp1}) and then prove feasibility 
of the program (\ref{pollp1}). By (\ref{pt0}) we obtain
\begin{equation}
Z_t(0) = {\gamma^{2t}h(x^\ast)  
\over aK_t^2(q;x^{\ast})}{n \choose t}^2 
\Big( {t+1+\gamma(n-t) 
\over t+1}\Big) ^2 \label{Z(0)}
\end{equation}
Further, by \cite{aal77}, 
\begin{equation}
\label{z0}
z_0={q\gamma^t h(x^\ast) 
\over (t+1)K_t^2(q;x^{\ast}) }{n \choose t}.
\end{equation}
We would like to substitute these values in (\ref{eq:main-bound}).
Recall the notation $R^\bot=(1/n)\log_q|C^\bot|=(1/2)(1+R_Q).$
Note that whenever 
\[
{\log_q{n\choose t}^2\gamma^{2t}\over 
\log_q \Big({n \choose t}\gamma^t |C^\bot|\Big)}\to0,
\]
i.e., $H_q(\tau_0(\xi^\ast))+R^\bot> 2H_q(\tau_0(\xi^\ast))$ or
\[
R^\bot\ge H_q(\tau_0(\xi^\ast))
\]
we have $Z_t(0)=o\Big( z_0 |C^\bot| \Big)$.
The restriction $R^\bot> H_q(\tau_0(\xi^\ast))$ by our choice of $\tau$ is 
equivalent to 
$R^\bot\ge R_q^{lp}(\xi^{\ast})$, which is always the case 
whenever $\xi^{\ast}\ge \delta_q^{lp}(R^\bot)$.
In this case the main term of the estimate  (\ref{eq:main-bound}) 
is given by the exponent of $ z_0 |C^\bot|.$
Differentiating $(1/n)\log_q z_0$ on $\xi^{\ast},$ we obtain
\[
\log_q{p(1-\xi^{\ast})\over (1-p)\xi^{\ast}}.
\]
The zero of this expression is $\xi^{\ast}=p,$ and 
$d(\log_q z_0)/d\xi^\ast$ it is negative for $\xi^{\ast}\in (p,1)$.  
Thus, the optimal choice of $\xi^{\ast}$ is $\xi^{\ast}=p$ if this value 
is not less than $\delta_q^{lp}(R^\bot )$ and 
$\xi^{\ast}=\delta_q^{lp}(R^\bot)$ otherwise. 
By (\ref{z0}), (\ref{krav_orthog}), and the fact that the
exponent of $h(x)$ is given by $-T_q(\xi,p)$, we obtain 
\begin{equation}\label{logz0}
{1 \over n}\log_q z_0= H_q(\tau) -T_q(\xi^\ast,p)-1 
-H_q(\tau)+H_q(\xi^{\ast})
\end{equation}
Substituting this in (\ref{eq:main-bound}), we obtain (\ref{lp1}).

Let us prove that polynomial (\ref{pollp1}) is admissible with respect to  
the restrictions (\ref{eq:ii})-(\ref{eq:iii}). 
\remove{The first one of them follows from the corresponding result
in \cite{aal77} since $Z_t(x)$ differs from $f_t(x)$ only by a (positive)
constant factor.}
The proof will be broken into two cases,\\[1mm]
(a) $\xi\in(0,\xi^{\ast}]$ and\\[1mm] 
(b) $\xi\in(\xi^{\ast},1)$.\\[1mm]
We begin with the first case and (\ref{eq:ii}).
We are only going to prove that it holds asymptotically,
i.e., to prove the inequality 
\begin{equation}
{1 \over n} \log_q Z_t(\xi n) \le  {1 \over n} 
\log_q h(\xi n). \label{maincon} 
\end{equation}
Here we employ a method used in the corresponding
part of \cite{ab}. Namely, by our choice of $\tau,$ the
smallest zero (\ref{root}) of $K_t$ tends to $\xi^\ast;$ 
hence in the interval
considered the exponent of $K_t$ is given by (\ref{logpt}).
Then we can write
\begin{align*}
{1\over n}&\log_qZ(\xi n)=-T_q(\xi^\ast,p)\\
&-2\int_{\xi^\ast}^\xi\log_q{(1-y)\gamma+y-q\tau+
\sqrt{((1-y)\gamma+y-q\tau)^2-4\gamma y(1-y)}\over 2\gamma(1-y)}
dy.
\end{align*}
with $o(1)$ terms omitted. Let $\psi(\tau,\xi):=2\int_0^\xi 
\dots dy+T_q(\xi,p);$
then we have 
\[
{1\over n}\log_qZ(\xi n)+T_q(\xi,p)=\psi(\tau,\xi)-\psi(\tau,\xi^\ast).
\]
Since $(1/n) \log_q Z_t(x^{\ast})=(1/n) 
\log_q h(x^{\ast}),$ all we need to prove is that
\begin{align*}
\psi'_\xi(\tau,\xi)=&2\log_q {(1-\xi)\gamma+\xi-q\tau+
\sqrt{((1-\xi)\gamma+\xi -q\tau)^2-4\gamma \xi(1-\xi)}\over 2
\gamma(1-\xi)}\\
&-\log_q{p\over(1-p)\gamma}>0, \quad 0\le\xi<\xi^\ast.
\end{align*}
First note that $\psi'_\xi(\tau,\xi)$ is a monotone decreasing
function of $\tau=\tau_0(\xi^\ast)$ and $\tau_0(z)$ is a monotone
decreasing function of $z$ (\ref{eq:tau}). Hence if we prove that
$\psi'_\xi$ is positive for $\xi^\ast=p$ this will also imply
that it is positive for $\xi^\ast=\delta_q^{lp}(R^\bot)>p.$
Therefore, let $\xi^\ast=p.$
In \cite[Appendix B]{ab} a similar function was proved to be positive.
The proof proceeds as follows: consider the difference
\[
g(\tau,\xi)=
{\big[(1-\xi)\gamma+\xi-q\tau+
\sqrt{((1-\xi)\gamma+\xi -q\tau)^2-4\gamma \xi(1-\xi)}\,\big]^2\over 4
\gamma^2(1-\xi)^2}-{p\over(1-p)\gamma},
\]
$\xi\in[0,\xi^\ast), \tau=\tau_0(p).$ The required inequality 
$\psi'_\xi(\tau_0(p),\xi)>0$
is implied by $g(\tau_0(p),\xi)\ge 0;$ the latter follows by the fact that
$g(\tau_0(p),p)=0$ and that upon substituting $\tau$ and simplifying
we obtain a fraction whose denominator is positive and the
derivative of the numerator on $\xi$ is negative in the whole segment 
$\xi\in[0,\xi^\ast].$

Now let us prove (\ref{eq:iii}) in case (a). 
According to \cite{al} the function 
${1 \over n} [\log Z_t(\xi n) - \log z_{\xi n}],  
\,\xi\in (0,\xi^{\ast}),$  achieves its maximum at $\xi=0$ for  
$1 \ge R_Q \ge \alpha_1(q)$\footnote{This is the reason for the
bound to fail for very low rates both in \cite{al} and here: 
the maximum shifts away from 0 and the analysis becomes unmanageable.}. 
If $p\le \delta_{LP1}(R^\bot)$ then as said above, we put
$\xi^{\ast}=\delta_{LP1}(R^\bot)$. This means that
\[ 
\log_q {Z_t(0)\over z_0}\to H_q\Big({t\over n}\Big)=
R_q^{lp}(\xi^\ast)=R^\bot,
\]
and so $z_0|C^\bot|-Z_t(0)\to 0.$ Therefore, for any
integer $s,\,  0< s \le x^{\ast},$ we have
$z_s|C^\bot|-Z_t(s)\ge 0.$

Finally if $p=\xi^\ast> \delta_q^{lp}(R^\bot)$
then  $(1/n)\log_q( Z_t(0)/z_0)$ converges to a number less than $R^\bot.$
Hence for sufficiently large $n$ and any integer $s\in (0,x^\ast]$
we have $z_s|C^\bot|-Z_t(s)>0.$ This takes care of case (a).

To verify feasibility in case (b), i.e., to prove
(\ref{eq:ii})-(\ref{eq:iii}) for $x^{\ast}< x \le n$,
we recall that $a$ is the smallest zero of $K_t(q;x)+K_{t+1}(q;x)$.
Let $y_s$ be the smallest zero of $K_s(q;x).$ Then by the well-known
properties of Krawtchouk polynomials we obtain that
$y_{t+1}< a < y_t$; so by (\ref{root}), $a\searrow y_{t+1}$ 
as $n \to \infty$. Then we have, for large $n$ and
all integer $x,\,x^{\ast}< x\le n,$
\[
Z_t(x)\le 0 \le {p^x \over \gamma^x}(1-p)^{n-x};
\]
hence (\ref{eq:ii}). To prove (\ref{eq:iii}), observe that if
$x > x^{\ast}$ then $Z_t(x)\le 0$ and so for any integer $s\ge x$
we have $z_{s}|C^\bot|-Z_t(s)\ge 0$.
This exhausts case (b) and completes
the proof.\qed

\subsection{Hamming-type bound}

In the $R$-$\delta$ problem for nonbinary codes
the bound \cite{aal77} is not the best one known.
It can be improved in several ways, in particular,
in the frame of the polynomial method a better result
is given in \cite{aal90}. However, the technique in \cite{aal90}
does not readily carry over to the present situation.
Another, somewhat simpler bound that improves upon \cite{aal77}
is the Hamming one which is better for a certain segment
of rates close to 1. Therefore, in this subsection 
we derive a Hamming-type bound on $E(R_Q,p).$ 
This bound is valid for low error probabilities: $p\in[0,p_{cr}],$
where the critical value $p_{cr}$ depends on $q$ (it is $0.19$ for
$q=2$ and $0.30$ for $p=4$). This improves
Theorem \ref{thm:qlp1} for some values of $R_Q$ dependent
on $q$ and further extends the segment in which the exponent
$E(R_Q,p)$ is known exactly.

We begin with the polynomial 
\[
F_e(x) =\sum_{i=0}^n f_i K_i(q;x),
\]
where $f_i=K_{e}(q;i)^2$. This polynomial is used in the
proof of the Hamming bound on the size of the code with a
given minimum distance \cite{del72}. Our first goal is to show how to modify
it for use in our problem. 

Delsarte \cite[p.13]{del73} proved that 
\[
K_i(q;x)K_j(q;x)=\sum_{k=0}^n p_{ij}^k K_k(q;x).
\]
where $p_{ij}^k$ are the intersection numbers of the $q$-ary 
Hamming scheme. By a straightforward generalization of the binary
case \cite[(A.19)]{ref mcel} we have
\[
 p_{ij}^k=\sum_{s=0}^{n-k} {k \choose 2k+2s-i-j } 
{n-k \choose s}
{2k+2s-i-j \choose k+s-j} (q-2)^{i+j-2s-k}(q-1)^s
\]
Therefore, 
\[
f_i=\sum_{k=0}^n p_{ii}^kK_k(q;i).
\]
Substituting in $F_e(x)$, we obtain the following: 
\begin{multline}
F_e(x) =\sum_{j=0}^n\sum_{k=0}^n\sum_{s=0}^{n-k}
{k\choose 2k+2s-2e}{n-k\choose s}{2k+2s-2e \choose k+s-e }\nonumber\\
\cdot(q-2)^{2e-2s-k}(q-1)^s
K_k(q;j)K_j(q;x)\nonumber\\
=\sum_{k=0}^n\sum_{s=0}^{n-k}
{k\choose 2k+2s-2e}{n-k\choose s}{2k+2s-2e \choose k+s-e }\nonumber\\
\cdot(q-2)^{2e-2s-k}(q-1)^s
\sum_{j=0}^{n} K_k(q;j)K_j(q;x)\nonumber\\
=q^n\sum_{s=\max\{0,e-x\}}^{e-x/2}
{x\choose 2x+2s-2e}{n-x\choose s}{2x+2s-2e \choose x+s-e 
}(q-2)^{2e-2s-x}(q-1)^s, \label{eq:F_e(x)}
\end{multline}
where in the last step we made use of (\ref{orth1}). 

Let us analyze the asymptotics of $F_e(x).$ Letting
$x=\xi n, s=\sigma n,$ and $e= \tau n$, we can write the
exponent of the summation term as follows:
\begin{multline}
\frac{1}{n}
\log_q \Big[{x\choose 2x+2s-2e}{n-x\choose s}{2x+2s-2e \choose x+s-e 
}(q-2)^{2e-2s-x}(q-1)^s\Big]\\
=\Big[\xi H_2\Big(\frac{2\xi+2\sigma-2\tau}{\xi}\Big)
+2\xi+2\sigma-2\tau\Big]\log_q2+
(1-\xi)H_q\Big(\frac{\sigma}{1-\xi}\Big)\\
+(2\tau-2\sigma-\xi)\log_q (q-2)+O\Big(\frac{1}{n}\Big).
\label{eq:exponent}
\end{multline}
Computing the derivative of the last expression on $\sigma$
and equating it to $0$, we arrive at the following condition:
\begin{equation}
\label{alpha}
{(2\sigma+\xi-2\tau)^2(1-\xi-\sigma)(q-1)\over
\sigma(2\sigma+2\xi-2\tau)^2(q-2)^2}={1\over 4}.
\end{equation}
It is not difficult to check (see \cite[Appendix]{al}) that this 
equation has only one root in the interval $\max\{0,\tau-\xi \} < 
\sigma < \tau-\xi/2 $. Denote this root by $\sigma_0.$
The main term in $F_e(x)$ asymptotically corresponds to the
value $s=\lfloor\sigma_0 n\rfloor.$
Thus, defining 
\begin{multline}\label{eq:phi}
\phi(\tau,\sigma,\xi):=1+(1-\xi)H_q\Big(\frac{\sigma}{1-\xi}\Big)+
\Big[\xi H_2\Big(\frac{2\xi+2\sigma-2\tau}{\xi}\Big)+2\xi+
2\sigma-2\tau\Big]\log_q2\\
+(2\tau-2\sigma-\xi)\log_q (q-2),
\end{multline}
we observe that
\[
\frac{1}{n}\log_q F_e(x)= \phi(\tau,\sigma_0,\xi)+O\Big(\frac{1}{n}\Big).
\]
The analysis is complicated by the fact that 
$\sigma_0$ itself is a function of $\tau$ and $\xi$.

Our general plan is, starting with $F_e(x)$, to construct a
polynomial $Z(x)$ so that $Z(x)$ be equal to $h(x)$ at one point and
less than $h(x)$ at all other integer points of the interval, thus
guaranteeing feasibility. Together with (\ref{alpha})
this gives two conditions on the 2 parameters,
$\sigma_0$ and $\tau,$ both functions of $\xi.$
It remains to make a suitable choice for $\xi;$ this we simply 
guess prompted by an analogy in the binary case.
This is the actual sequence of steps that we perform to derive the
bound. Calculations, though elementary, are fairly involved,
and we will not write them out in full. Instead, we perform
a similar analysis in the binary case; this can be done
explicitly within reasonable space and fixes ideas for the 
general result.

\medskip{\sc the binary case}.
We have the following simple expression for $F_e(x)$:
\begin{align}
F_e(x)&=2^n{x \choose x/2}{n-x \choose e-x/2}\nonumber \\
&=\sum_{i=0}^nK_i(2,x)\sum_{k=0}^n{n-k\choose e-k/2}{k\choose k/2}K_k(2;i). 
\label{eq:F_e(x)bin}
\end{align}
Note that $F_e(x)=0$ when $x>2e$. The exponents of $F_e(x)$ and $h(x)$ are
\begin{align}
\label{eq:phibin}
\phi(\tau,\xi):={1 \over n}\log_2 F_e(x)&=1+\xi+(1-\xi)H_2
\Big({2\tau-\xi \over 2-2\xi}\Big)+o(1),\\ 
{1 \over n}\log_2 h(x)&=-T_2(\xi,p). \nonumber
\end{align}
Now let $x^\ast=\xi^\ast n$ be a point at which these exponents have equal
slopes. Let us first convince ourselves 
that such a point exists and is unique. Indeed, the polynomial
$\phi_\xi'(\tau,\xi)+(T_2(\xi,p))'_\xi$ is quadratic in $\xi;$
its zeros are
\[
1\pm{(1-p)(1-2\tau)\over\sqrt{1-2p}}.
\]
Of them the one with the $+$ sign is greater than one; the 
other one is always between $0$ and $1.$ 

The equation $\big[\phi_\xi'(\tau,\xi)=-(T_2(\xi,p))'_\xi\big]_{\xi=\xi^\ast}$
defines $\tau$ as a function of $\xi^\ast.$
Namely,
\begin{equation}
\label{eq:tau-ast}
\tau=\tau(p)={1\over2}\Big(1-{\sqrt{1-2p}\over1-p}(1-\xi^\ast)\Big).
\end{equation}
\remove{

Derivatives of exponents of $F_e(x)$ and $h(x)$ are
\begin{equation}
{d \over d\xi} \Big({1 \over n} \log F_e(\xi n)\Big)=1-H\Big({2\tau-\xi 
\over 2-2\xi}\Big)+{(2\tau-1) 
\over 2(1-\xi)}\log\Big({2-\xi-2\tau \over 2\tau-\xi}\Big),
\end{equation}
\begin{equation}
{d \over d\xi} \Big({1 \over n} \log h(\xi n)\Big) =\log {p\over 1-p}.
\end{equation}
Given $\tau$ and $p$ we can find $\xi^{\ast}$ such that 
these derivatives are equal 
to each other.
After simple computations, we get that

Note that this function can be easily inverted
\begin{equation}
\label{tauinv}
\tau=\tau(\xi^{\ast}):={1\over 2}+{1\over
2}{\sqrt{1-2p}(\xi^{\ast}-1) 
\over 1-p}
\end{equation}
This gives us the value of $e=\tau n$ to use in $F_e(x)$

}

Now we can define the polynomial $Z(x)$ by rescaling $F_e(x)$ as
follows. Let
\[
Z(x)={h(x^\ast)\over  F_{e}(x^\ast)} F_{e}(x),
\]
where $e=\lfloor \tau n\rfloor.$
Note that we have ensured
that $Z(x)$ equals $h(x)$ at $x^\ast$ and that their exponents
are tangent; it will be seen below
that for large $n,$ $Z(x)<h(x)$ at all the other integer
points of the interval. It remains to choose the value of $\xi^\ast$.
This point is taken to maximize $z_0$, i.e., 
the estimate (\ref{eq:main-bound}). 
Namely, the logarithm of $z_0$ equals 
\begin{equation}
\label{hamz0}
{1 \over n}\log_2 z_0=2H_2(\tau)-T_2(\xi^{\ast},p)-\phi(\tau,\xi^\ast)+o(1).
\end{equation}
Substituting $\tau$ from (\ref{eq:tau-ast}) and
taking the derivative on $\xi^\ast$, we find that
the optimal choice is $\xi^\ast=p.$
\remove{
\begin{align*}
{d \over d\tau } \Big({1 \over n} \log_2 z_0 \Big)  =&-{(1-p) \over 
\sqrt{1-2p}}\log_2
\Big( {(1-p)^2-(1-2p) \over p^2}\Big)\\
&-\log_2\Big(
{1-p+\sqrt{1-2p} \over 1-p-\sqrt{1-2p}}\Big)
-2\log_2 {\tau \over 1-\tau};
\end{align*}
equating this to $0$ we find
\begin{equation}
\label{tst}
\tau={1\over 2}-{1\over 2} \sqrt{1-2p}.
\end{equation}
}
Note that substituting $\xi^\ast=p$ and $\tau$ from (\ref{eq:tau-ast})
in (\ref{hamz0}), we find $(1/n)\log_2 z_0\to-1.$ 

Let us examine feasibility of $Z(x).$ As a preliminary remark, note that
we are allowed to put
$\xi^\ast=p$ as long as $p$ is greater than the Hamming
distance and we have to take $\xi^\ast$ equal to this distance otherwise.
Indeed, substituting $x=0$ in
(\ref{eq:F_e(x)bin}), we obtain
\begin{gather*}
{1 \over n}\log_2 F_{\tau n}(0)\sim 1+H_2(\tau) \\
{1 \over n}\log_2 f_0={1\over n}\log K_{\lfloor\tau n\rfloor}^2(2;0)
\sim 2H_2(\tau),
\end{gather*}
the latter by (\ref{pt0}). From this and the definition of $Z(x)$ 
it follows that whenever 
\begin{equation}
\label{hamcondbin}
H_2(\tau)\ge 1-R^\bot,
\end{equation}
we have $Z(0)=o(z_0|C^\bot|)$, needed for the estimate (\ref{eq:main-bound})
to be nontrivial. So we can choose $\xi^{\ast}=p,\, \tau=\tau(p)$
if (\ref{hamcondbin}) holds and we choose $\tau=H_2^{-1}(1-R^\bot)$ 
(the Hamming distance for the rate $R^\bot$) and $\xi^{\ast}$ the root of
$H_2(\tau(\xi))=1-R^\bot$ otherwise.

Note that by (\ref{eq:tau-ast}) $\tau$ grows on $\xi^\ast$.
Thus, (\ref{eq:ii}) will follow in both cases
if we prove that it holds for all $(1/2)(1-\sqrt{1-2p})\le\tau\le (1/2).$
As above, we are only going to prove that (\ref{eq:ii}) holds
asymptotically, i.e., that
\begin{equation}\label{eq:feasibility}
\log_2Z(x)-\log_2h(x)\le 0, \quad 0\le\xi\le2\tau, \;
{1\over 2}(1-\sqrt{1-2p})\le\tau\le {1\over2}.
\end{equation}

We begin with the case $\xi^\ast=p$. 
Note that $(1/n)\log_2h(x)$ is a straight line; its derivative is
$\log_2(p/(1-p))<0.$
Inequality (\ref{eq:feasibility})
will follow from the following set of conditions:\\[2mm]
\begin{tabular}{ll}
(i) &$\phi'_\xi(\tau,\xi)=\log_2(p/(1-p))$ has a unique zero 
for $\xi\in [0,2e]$;\\[1mm]
(ii) &$0>\phi'_\xi(\tau,\xi)|_{\xi=0}>\log_2(p/(1-p));$\\[1mm]
(iii) &$\phi'_\xi(\tau,\xi)|_{\xi=2e}<\log_2(p/(1-p))<0$,\\[1mm]
\end{tabular}\\[2mm]
where $\tau=(1/2)(1-\sqrt{1-2p}).$

Condition (i) was established above.
Conditions (i)-(iii) imply that $\phi_\xi(\tau,\xi)<-T_2(\xi,p)$ 
for $\xi\in[0,p)$
and $\phi_\xi(\tau,\xi)<-T_2(\xi,p)$ for $\xi\in[0,p).$
Indeed, suppose that $\phi(\tau,\xi)>\log_2(p/(1-p)$
in the neighborhood of $\xi^\ast,$ say on the left of it. 
This implies that in the small neighborhood of the point of tangency
the derivative $\phi'_\xi$ is smaller that $\log_2(p/(1-p));$ however by (ii)
it is greater than that for $\xi=0$, hence there is another point between $0$ 
and $p$ at which $\phi$ and $-T_2$ have equal slopes, but this
violates (i). Supposing that $\phi$ and $-T_2$ intersect at some
point between $0$ and $\xi^\ast$, we again find a similar contradiction.
The second part of the claim follows by the same
argument. Thus to establish (\ref{eq:feasibility}) 
it suffices to prove (ii)-(iii).
\remove{To prove (i) for large $n$,
let us bound above the exponent of $Z(0)/h(0),$ which is
\begin{equation}\label{eq:comp}
T_2(0,p)-T_2(\xi^\ast,p)-\phi(\tau,\xi^\ast)+\phi(\tau,0),
\end{equation}
or, substituting $\xi^\ast=p$ and $\tau=(1/2)(1-\sqrt{1-2p})$ and
expanding everything, the quantity
\begin{gather*}
\log_2{1\over
1-p}-H_2(p)-p-(1-p)H_2\Big({1-\sqrt{1-2p}-p\over2-2p}\Big)\\
+H_2\Big({1\over 2}(1-\sqrt{1-2p})\Big).
\end{gather*}
This function is zero for $p=0,(1/2)$ and negative everywhere in between.
This proves (i).
}

We have
\[
\phi'_\xi(\tau,\xi)={1\over2}\log_2{(2\tau-\xi)(2-2\tau-\xi)\over(1-\xi)^2};
\]
so 
\[
\phi'_\xi((1/2)(1-\sqrt{1-2p},0)={1\over2}\log_22p>\log_2{p\over 1-p}
\quad(0\le p<{1\over2}).
\] 
This proves (ii).
Condition (iii) is equally elementary; we omit the easy check.
This establishes (\ref{eq:ii}) for $\xi^\ast=p.$

Now suppose that $\xi^\ast>p.$ Condition (i) was proved above for any
$\tau.$ To prove (ii) and (iii) we only have to
show that $\phi'_\xi(\tau,0)$ grows and $\phi'_\xi(\tau,2\tau)$
falls on $\tau.$ Observe that 
$\phi'_\xi(\tau,0)={1\over2}\log_24\tau(1-\tau)$ indeed grows as
long as $\tau<1/2,$ which is true, and $\phi'_\xi(\tau,\xi)$ falls 
indefinitely as $\xi\to 2\tau$. This proves (\ref{eq:ii}), or rather
(\ref{eq:feasibility}), for $q=2.$

To prove (\ref{eq:iii}), we choose the following tactics. We already 
know by (\ref{hamcondbin}) that (\ref{eq:iii}) holds for $\xi\to 0.$
Hence it suffices to prove that the expression 
$\log_2 Z(\lfloor\xi n\rfloor)/z_{\lfloor \xi n \rfloor}$ 
achieves its maximum at $\xi =0$. 
Numerical computations show that this condition holds if 
$0\le \tau \le \tau_1\approx 0.1069$. 

Finally, by the definition, $Z(x)=0$ for $x\ge \lceil 2\tau n\rceil$.
Hence for these values of $x$ (\ref{eq:iii}) holds trivially.
Otherwise by the preceding paragraph, (\ref{eq:iii}) is true at least as
long as $i$  is less than the smallest root of $K_{\lfloor \tau n\rfloor}$
since otherwise the coefficients $z_i=(K_{\lfloor \tau n\rfloor}(i))^2$ 
of $Z(x)$ can be very small. The smallest zero is given by
(\ref{eq:tau}); so the discussed
constraint is satisfied in particular if $\tau\le \tau_2,$ where
$\tau_2=0.1$ is a root of 
$2\tau=(1/ 2)-\sqrt{\tau(1-\tau)}$.
Thus a sufficient condition for (\ref{eq:iii}) to hold true is that
$\tau\le \tau_{cr}=\min\{\tau_1,\tau_2\}=0.1$.
Note also that $\tau(p)$ is monotone increasing in $p$. Hence
we substitute $\xi^\ast=p$ in (\ref{eq:tau-ast}) and denote
by $p_{cr}$ the root of $\tau(p)=\tau_{cr}$, $p_{cr}=0.18.$
 
In summary, we obtain the following theorem. 

\begin{thm} \label{thm:qhambin} 
Let $R=(1/2)(1-R_Q).$ Then for any $0\le p\le p_{cr}$ 
\begin{equation*}
E(R_Q,p)\le \begin{cases}
-1-R+T_2(\xi,p)+\phi(H_2^{-1}(R),\xi) &1-2H_2(\tau_{cr})
\le R_Q\le 1-2H_2(\tau(p));\\
R &1-2H_2(\tau(p))\le R_Q\le 1,
\end{cases}
\end{equation*}
where $\xi$ is a root of $H_2(\tau(\xi))=R,$ and
$\phi(\cdot)$ is given by {\rm(\ref{eq:phibin})}.
\end{thm}

This completes the argument in the binary case.

\medskip{\sc the general case.} The analysis is similar but 
significantly more complicated since apart from $\tau$ and $\xi$ 
we have a third parameter, $\sigma$. In this part we are more sketchy 
than above.
The polynomial $Z(x)$ is again taken in the form
\[
Z(x)={h(x^\ast)\over F_e(x^\ast)}F_e(x),
\]
where this time $h(x)$ is as in (\ref{eq:h(x)}) and $x^\ast=\xi^\ast n$
is a parameter.

By (\ref{eq:phi}) 
\begin{gather*}
(1/n)\log_q
Z(x)=\phi(\tau,\sigma_0,\xi)-T_q(\xi^\ast,p)-
\phi(\tau,\sigma_0,\xi^\ast).
\end{gather*}
We proceed exactly as above. Namely, from the equation 
$\phi'_\xi(\tau,\sigma_0,\xi^\ast)=(T_q(\xi^\ast,p))'_\xi$ 
we find $\tau$ as a function of $\xi^\ast$ and $\sigma_0.$ This gives
\begin{multline*}
\tau= 
\tau(\xi^{\ast},\sigma_0):=
{\xi^\ast(\xi^\ast+\sigma_0-1)(q-1-p)-p\sigma_0(q-2)
\over p(\xi^\ast-1)(q-2)}\\
+{\sqrt{(q-1)(1-p)\xi^{\ast 2}\big[(1-\xi^\ast)^2(q-1-p)
-\sigma_0(1-\xi^\ast)(2q-2-qp)+\sigma_0^2(q-1)(1-p)
\big]}\over 
p(\xi^\ast-1)(q-2)}.
\end{multline*} 
Next, we substitute this value of $\tau$ in (\ref{alpha}) and 
find $\sigma_0$ as a function of $\xi^\ast.$ This gives
\begin{equation}\label{eq:sigma0}
\sigma_0:=\sigma_0(\xi^\ast)={2q-2-qp-2(1-\xi^{\ast})\sqrt{(q-1)(q-1-qp)}
\over q^2(1-p)}.  
\end{equation}
To complete the definition of the parameters we have to chose
$\xi^\ast.$ 
As above, we take $\xi^\ast=p$ as long as this does not violate
the feasibility condition (\ref{eq:iii})\footnote{Though we do not
prove this, this choice of $\xi^\ast$ is optimal with respect to the
bound (\ref{eq:main-bound}).}. Substituting $x=0$ in
(\ref{eq:F_e(x)}), we obtain
\begin{gather*}
{1 \over n}\log_q F_{\tau n}(0)\sim 1+H_q(\tau) \\
{1 \over n}\log_q f_0={1\over n}\log K_{\lfloor\tau n\rfloor}^2(q;0)
\sim 2H_q(\tau),
\end{gather*}
the latter by (\ref{pt0}). From this and the definition of $Z(x)$ 
it follows that whenever 
\begin{equation}
\label{hamcond}
H_q(\tau)\ge 1-R^\bot
\end{equation}
we have $Z(0)=o(z_0|C^\bot|)$. 

So the best possible choice is $\xi^{\ast}=p,\,  \tau=\tau(p,\sigma_0)$
if (\ref{hamcond}) holds  
and $\tau=H_q^{-1}(1-R^\bot)$ (the Hamming distance
for the rate $R^\bot$) and $\xi^{\ast}$ the root of
$H_q(\tau(\xi,\sigma_0))=1-R^\bot$ otherwise. Computations with 
Maple show that in the first case
${1\over n}\log_q z_0\to -1,$ exactly as in the binary case above. 
We did not find a closed-form expression for the second 
case.   

Similarly to the binary case $\tau$ should satisfy the inequality
\[
2\tau\le {\gamma \over 
q}-{\gamma-1 \over q}\tau
-{2 \over q}\sqrt{\gamma\tau(1-\tau)} \quad\mbox{ (cf. (\ref{eq:tau}))},
\]
or
$$
\tau\le \tau_2:={(3\sqrt q-2\sqrt2)(q-1)\over\sqrt q(9q-8)}.
$$
Also similarly to the binary case we have to choose $\tau$ such that 
\begin{equation}\label{maximum}
\arg\max_\xi
\big\{\log_2 {Z(\lfloor \xi n\rfloor)\over z_{\lfloor \xi n\rfloor}}\big\}=0.
\end{equation}
For $\tau=0$ this maximum is obviously achieved at $\xi=0$ 
(note that $F_0(x)=q^n\delta_{x0}$ and $f_{\lfloor x\rfloor}=1$). 
Define $\tau_1, 0\le \tau_1\le \tau_2,$ as follows:
\[
\tau_1=\inf(\tau: \mbox{ (\ref{maximum}) does not hold}).  
\]
Let $\tau_{cr}=\tau_1$ if $\tau_1$ is well-defined and 
$\tau_{cr}=\tau_2$ otherwise. The function again 
$\tau(p)$ is increasing in $p$. Let $p_{cr}$ be the root of 
$\tau(p,\sigma_0(p))=\tau_{cr}$. 
Now we are ready to formulate the theorem.

\begin{thm}
\label{thm:qham} Let $R=(1/2)(1-R_Q).$ Then for any $0\le p\le p_{cr}$ 
\begin{equation*}
E(R_Q,p)\le\left\{ \begin{array}{ll}
-1-R+&T_q(\xi^\ast,p)+\phi(H_q^{-1}(R),\sigma_0(\xi^\ast),\xi^\ast),\\ 
&1-2H_q(\tau_{cr}) \le R_Q\le 1-2H_q(\tau(p,\sigma_0));\\[2mm]
R, &1-2H_q(\tau(p,\sigma_0))\le R_Q\le 1,  
\end{array}
\right.
\end{equation*}
where $\xi^\ast$ is a root of $H_q(\tau(\xi,\sigma_0(\xi)))=R,$
$\sigma_0$ is defined in {\rm(\ref{eq:sigma0})}, and
$\phi(\cdot)$ is given by {\rm(\ref{eq:phi})}.
\end{thm}

{\bf Remark} For $q=4,$ $p_{cr}=0.301,$ and numerical computations 
show that (\ref{maximum}) holds true in the entire interval 
$\tau\in[0,\tau_2].$ Therefore in this case $\tau_{cr}=\tau_2$.  

Fore reference purposes we composed a short table of values of
the bounds for $q=4,\,p=0.1.$

\medskip
\begin{tabular}{|l|l|l|l|}
\hline
$R_Q$ & Existence & A-MRRW & Hamming \\
\hline
0& 0.5260& 0.6270&  -- \\
0.1 &  0.4637& 0.5458&  -- \\
0.2& 0.4054& 0.4685& 0.4774\\
0.3& 0.3509& 0.3952& 0.3951\\
0.4& 0.3& 0.3262& 0.3216\\
0.5& 0.25& 0.2618& 0.2567\\
0.6& 0.2& 0.2028& 0.2003\\
0.7& 0.15& 0.15& 0.15 \\
0.8& 0.1& 0.1& 0.1 \\
0.9& 0.05& 0.05& 0.05 \\
1  & 0 & 0 & 0\\
\hline
\end{tabular}

\medskip
These bounds are also plotted in Fig.\,1. It can be seen that the
Hamming bound is the best of the two upper bounds for large rates.
Unlike the classical case, the upper bounds do not approach the
lower bound as the rate $R_Q$ becomes small.
However this is due rather to the way of measuring the
rate of quantum codes than to an imperfection of the method.
Indeed, roughly speaking, the case $R_Q=0$ corresponds 
to classical codes of rate $R=1/2$ (cf. (\ref{eq:r})). 
The function $E(R_Q,p)$ is known exactly at least for
$2R_4^{lp}(p)-1\le R_Q\le 1.$ In fact, by Theorem \ref{thm:qham}
the left end of this interval is provably smaller than this value;
however, it is difficult to make any exact statements other than just
plotting the bounds.

\section{Appendix}
Let $K_i(q;x)$ be the $q$-ary Krawtchouk polynomial, $\gamma=q-1.$
Here we list its properties used in the paper.

The following 3 basic facts are well known:
\begin{gather}
\sum_{i=0}^n K_i(q;x)z^i=(1+\gamma z)^{n-x}(1-z)^x \quad
\mbox{($x$ integer)};\label{gen_fun}\\
\sum_{i=0}^n K_r(q;i)K_i(q;s)=q^n \delta_{rs};
\label{orth1}\\
f(x)=\sum_{i=0}^t f_i K_i(q;x) \quad\Leftrightarrow\quad
f_i=q^{-n} \sum_{j=0}^n f(j) K_j(q;i) ,\label{coef}
\end{gather}
where in (\ref{coef}) $f(x)$ is any polynomial with $\deg f\le n.$

Let $y_s$ be the smallest zero of $K_s.$ For $s=\sigma n,$  $n\to\infty$
we have \cite{aal77},
\begin{equation}
\label{root}
\frac{y_s}{n} =\tau(\sigma)+ o(1),
\end{equation}
where the function $\tau(\cdot)$ is defined in (\ref{eq:tau}).
Further, by \cite{kala95} for $n\to\infty$ and $\xi\in[0,\tau(\sigma)]$ 
\begin{gather} 
{1 \over n} \log K_s(q;\xi n) \sim H_q(\sigma) \nonumber \\
+\int_{0}^{\xi} 
\log {(1-y)\gamma+y-q\sigma +\sqrt{ ((1-y)\gamma+y-q\sigma)^2-4\gamma y 
(1-y)} \over 2\gamma (1-y) } dy.\label{logpt}
\end{gather}
In particular, for $\sigma=\tau(\xi),$ i.e., $\xi=\tau(\sigma),$
\begin{equation}  
\label{krav_orthog}  
{1 \over n} \log_q K_{\sigma n}(q;\xi n)={1+H_q(\sigma)-H_q(\xi)\over 2} 
+o(1).\end{equation} 
Finally, from the definition of $K_s$ we find
\begin{equation}
\label{pt0}
K_s(q;0)={n \choose s}\gamma^{s}.
\end{equation}

\footnotesize{

\begin{figure}
\centerline{\psfig{file=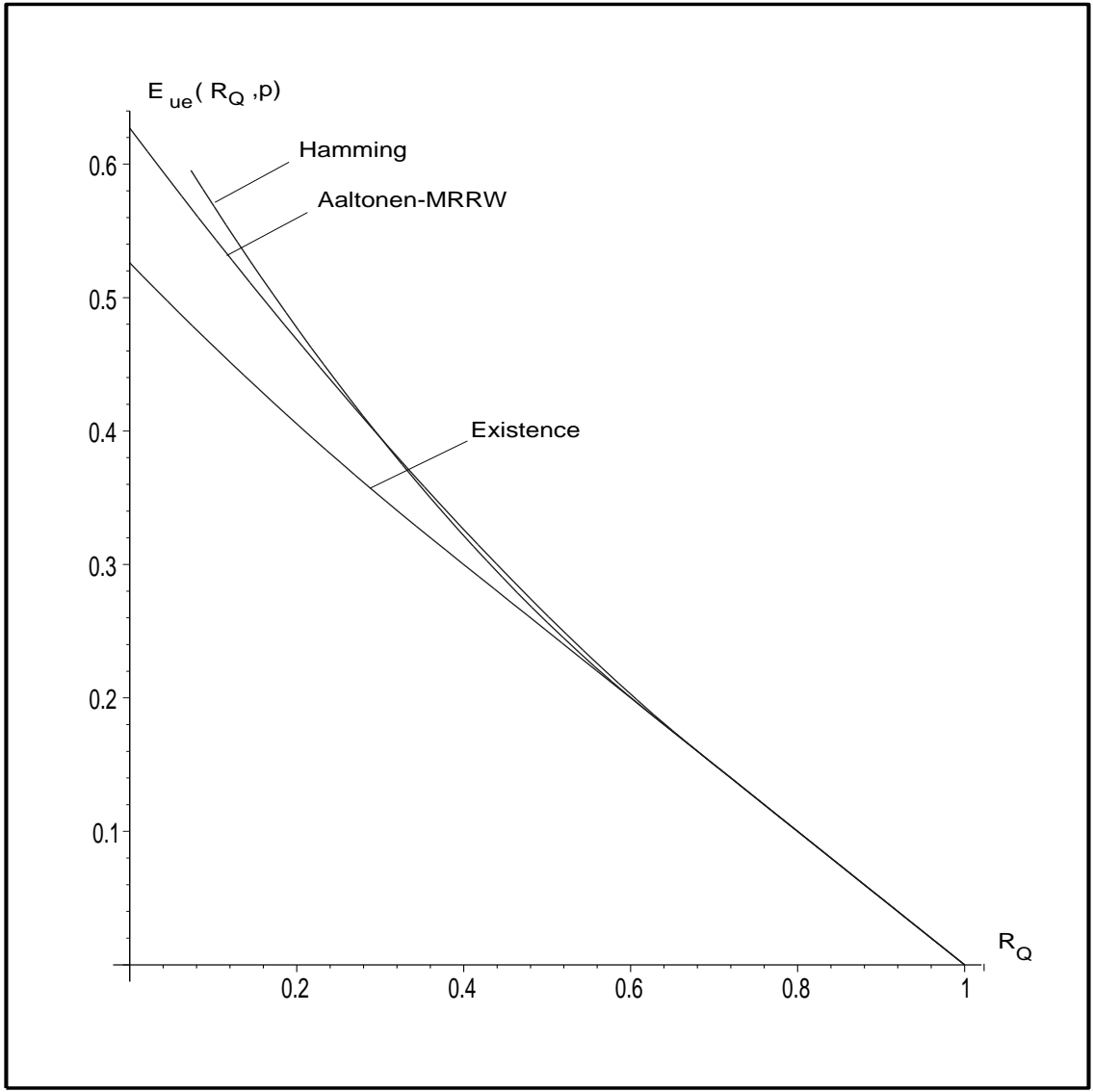,height=15cm,width=15cm}}
\caption{Bounds on $E(R_Q,p)\!:$ Existence bound, Theorem \ref{GV};
Aaltonen-MRRW-type bound, Theorem \ref{thm:qlp1};
Hamming-type bound, Theorem \ref{thm:qham}; $p=0.1.$}
\label{ref.fig1}
\end{figure}

\end{document}